\documentclass[twocolumn]{aastex61}

\usepackage[T1]{fontenc}
\usepackage{rotating}
\usepackage[figure,figure*]{hypcap}
\received{}
\revised{}
\accepted{}
\submitjournal{ApJ}

\shorttitle{GX 339--4 QPO phase lags}
\shortauthors{Zhang et al.}

\begin{document}

\title{The Evolution of the Phase Lags Associated with the Type-C Quasi-periodic Oscillation in GX 339--4 during the 2006/2007 Outburst}

\author{Liang Zhang}
\affiliation{Department of Astronomy, Beijing Normal University, Beijing 100875, China}
\affiliation{Kapteyn Astronomical Institute, University of Groningen, PO Box 800, NL-9700 AV Groningen, the Netherlands}

\author{Yanan Wang}
\affiliation{Kapteyn Astronomical Institute, University of Groningen, PO Box 800, NL-9700 AV Groningen, the Netherlands}

\author{Mariano M\'{e}ndez}
\affiliation{Kapteyn Astronomical Institute, University of Groningen, PO Box 800, NL-9700 AV Groningen, the Netherlands}

\author{Li Chen}
\affiliation{Department of Astronomy, Beijing Normal University, Beijing 100875, China}

\author{Jinlu Qu}
\affiliation{Key Laboratory for Particle Astrophysics, Institute of High Energy Physics, CAS, Beijing 100049, China}

\author{Diego Altamirano}
\affiliation{Department of Physics and Astronomy, University of Southampton, Southampton, Hampshire SO17 1BJ, UK}

\author{Tomaso Belloni}
\affiliation{INAF-Osservatorio Astronomico di Brera, via E. Bianchi 46, I-23807 Merate, Italy}

\correspondingauthor{Liang Zhang}
\email{201431160006@mail.bnu.edu.cn}

\begin{abstract}
  We present the evolution of the phase lags associated with the type-C QPO in GX 339--4 during the rising phase of the 2006/2007 outburst. We find that the phase lags at the QPO frequency are always positive (hard), and show very different behavior between QPOs with frequencies below and above $\sim1.7$ Hz: when the QPO frequency is below $\sim1.7$ Hz, the phase lags increase both with QPO frequency and energy, while when the QPO frequency is above $\sim1.7$ Hz, the phase lags remain more or less constant. When the QPO frequency is higher than $\sim1.7$ Hz, a broad feature is always present in the lag-energy spectra at around 6.5 keV, suggesting that the reflection component may have a significant contribution to the phase lags. Below $\sim1.7$ Hz, the QPO rms first decreases with energy and then turns to almost flat, while above $\sim1.7$ Hz, the QPO rms increases with energy. During the transition from the low-hard state to the hard-intermediate state, the second harmonic and subharmonic of this QPO appear in the power density spectra. The second-harmonic and subharmonic phase lags show very similar evolution with their centroid frequencies. However, the energy dependence of the second-harmonic and subharmonic phase lags are quite different. Our results suggest that, at different phases of the outburst, different mechanisms may be responsible for the phase lags of the QPO. We briefly discuss the possible scenarios for producing the lags.

\end{abstract}


\section{Introduction} \label{sec:intro}

Most of the known black-hole X-ray binaries are transient systems (Black Hole Transients, BHTs) that usually go through several distinct temporal and spectral states during a complete outburst. Following the classification of \citet{Homan2005}, these states are the low-hard state (LHS), hard-intermediate state (HIMS), soft-intermediate state (SIMS) and high-soft state (HSS) (see \citealt{Belloni2010,Belloni2011} for reviews; for a different classification scheme of the black-hole states, see \citealt{Remillard2006}). In the low-hard state, typically seen at the beginning and the end of an outburst, the energy spectrum is dominated by a hard power-law component ($\Gamma\sim1.5$) with a high-energy cutoff around $\sim100$ keV. The hard component is attributed to Compton up-scattering of soft photons by hot electrons in a corona \citep{Gilfanov2010}. In the high-soft state, usually observed in the middle of the outburst, the energy spectrum is dominated by a thermal component associated with an optically-thick, geometrically-thin accretion disc \citep{Shakura1973}. Two additional states, HIMS and SIMS, have been identified between the LHS and HSS \citep{Belloni2005}. During these states, the power-law component becomes softer than that in the LHS and the contribution from the disc component gradually increases.

The X-ray spectrum of BHTs also shows a reflection component, produced via the hard X-ray photons in the corona irradiating the inner part of the accretion disc \citep{Fabian2010}. The most prominent feature of the reflection spectrum is the fluorescent Fe K$\alpha$ line at around 6.4 keV, which is often broadened by Doppler effects, light bending and gravitational redshift \citep{Fabian1989}. Another important feature is the reflection hump between 10 and 30 keV due to Compton back-scattering (\citealt{Ross2005}; see \citealt{Miller2007} for a review).

Low-frequency quasi-periodic oscillations (QPOs) with centroid frequencies ranging from a few mHz to $\sim30$ Hz have been observed in the power density spectra (PDS) of most BHTs. These QPOs can be classified into three categories, dubbed type-A, B and C, based on the quality factor, fractional rms, noise component and phase lag properties \citep[e.g.,][]{Casella2005}. The energy dependence of the QPO amplitude shows that QPO emission is associated with the power-law component \citep{Gierlinski2005,Belloni2011,Gao2014}. Type-C QPOs are the most common type of QPOs in BHTs, and occur mostly in the LHS and HIMS. These QPOs are characterized by a strong, narrow peak with variable frequency, superposed on a band-limited noise continuum. A second harmonic and a subharmonic peak are sometimes present in the PDS. Here we will focus our study on the type-C QPO.

There is still no general agreement about the physical nature of the type-C QPOs, but increasing evidence suggests that they may have a geometric origin, with Lense-Thirring precession of the entire inner flow being the most promising model \citep[e.g.,][]{Ingram2009}. For instance, \citet{Heil2015} and \citet{Motta2015} found that the amplitude of the type-C QPO depends on the orbital inclination of the system in which they are observed. This confirms the prediction in the precessing ring model proposed by \citet{Schnittman2006} that the QPO should be stronger in high-inclination systems.

Based on phase-resolved spectroscopy analysis, the iron line equivalent width in GRS 1915+105 and the iron line centroid energy in H 1743--322 have been found to be modulated with QPO phase \citep{Ingram2015,Ingram2016}. Additionally, \citet{Ingram2017} found that the reflection fraction also changes with QPO phase. These results strongly point to a geometric QPO origin, and they are consistent with the idea that the QPO is produced by Lense-Thirring precession \citep{Stella1998,Schnittman2006,Ingram2009}.

Phase lags between different energy bands are a powerful tool to study the fast X-ray variability. Recently, \citet{Eijnden2017} made a systematic analysis of the phase lag in 15 BHTs and found that the phase lags at the type-C QPO frequency strongly depend on inclination. Low-inclination sources display hard lags (hard photons lag the soft ones) at high QPO frequencies, while high-inclination sources display soft lags, except GRS 1915+105 which shows both positive and negative lags. This is consistent with previous results from individual sources: XTE J1550--564 \citep{Remillard2002}, XTE J1859+226 \citep{Casella2004} and XTE J1752--223 \citep{Munoz2010}. Such an inclination dependence provides strong constraints on the physical mechanism that produces the phase lags. It is also very important to mention the peculiar source GRS 1915+105, which exhibits a very different phase-lag behavior. As QPO frequency increases, the phase lags decrease and change sign from positive to negative when the QPO is at around 2 Hz \citep{Reig2000}. Furthermore, the energy-dependent phase lags change with QPO frequency: for QPOs with a low frequency ($<2$ Hz), the phase lags increase with energy; for QPOs with a high frequency ($>2$ Hz), the phase lags decrease with energy \citep{Reig2000,Qu2010,Pahari2013}.

GX 339--4 is a recurrent low-mass X-ray binary (LMXB), which harbors a black hole with a mass function of $f(\rm{M})=5.8 \pm 0.5~M_{\odot}$ \citep{Hynes2003}. The distance to this system is between 6 and 15 kpc \citep{Hynes2004}. Due to the lack of eclipses and absorption dips \citep{Cowley2002}, the inclination of GX 339--4 must be less than $\sim60^{\circ}$, and a lower limit of $\sim40^{\circ}$ can be set by assuming that the black hole mass should not exceed $20~M_{\odot}$ \citep{Munoz2008}. GX 339--4 has undergone frequent outbursts and displayed all the black-hole states \citep[e.g.,][]{Mendez1997,Belloni1999,Belloni2005,Dunn2008} in the past thirty years, making it one of the most studied BHTs. Among these outbursts, the 2006/2007 outburst was the brightest one, and contains a large quantity of low-frequency QPO phenomena. Detailed studies on the evolution of the low-frequency QPOs and spectral parameters during this outburst have been carried out by \citet{Motta2009,Motta2011}. A systematic analysis of the reflection spectrum and the phase lags of the broad-band noise component in GX 339--4 has been done by \citet{Plant2014} and \citet{Altamirano2015}, respectively.

Tracing the evolution of rapid X-ray variability along an outburst can help us understand the physical changes of the accretion flow and the origin of the variability. Such crucial information can be obtained through studying the energy dependence of the variability properties, like QPO amplitude and phase lags. Therefore, in this paper we study in detail the evolution of the phase lags associated with type-C QPO in GX 339--4 during the rising phase of the 2006/2007 outburst. We measured these phase lags at different Fourier frequency ranges: QPO fundamental, second harmonic and subharmonic. We then produced a frequency-dependent phase lag spectrum for each observation and a energy-dependent phase lag spectrum for each Fourier component. In addition, we also calculated the energy-dependent QPO amplitude to investigate the QPO origin. We describe the observations and data analysis methods in Section 2. We present the results of our study in Section 3 and discuss them in Section 4. Conclusions follow in Section 5.

\section{Observations and data analysis} \label{sec:obs}

We analyzed 23 \emph{RXTE} public archival observations of GX 339--4 covering the rising phase of its 2006/2007 outburst. For our study, we selected only the observations that, according to \citet{Motta2009}, have shown a type-C QPO. Due to the low count rate and signal-to-noise ratio, it is hard to study the energy dependence of the QPO properties during the fall of the outburst. Therefore, we only considered the observations during the rise. Table \ref{tab:log} lists the log of \emph{RXTE} observations used in this work.

For the timing analysis, we used the software GHATS version 1.1.1 under IDL \footnote{GHATS, \url{http://www.brera.inaf.it/utenti/belloni/GHATS_Package/Home.html}}. For each observation, we computed an average PDS in the full energy band ($2-60$ keV). We used 128-s long intervals and 1/1024-s time resolution, corresponding to a Nyquist frequency of 512 Hz. The Leahy normalization was used \citep{Leahy1983} and the contribution due to the photon counting noise was subtracted \citep{Zhang1995}. We fitted the averaged PDS with a sum of Lorentzian functions using XSPEC v 12.9. In all PDS, a strong band-limited noise with one or more peaks were present, which is typical for the type-C QPO. Following \citet{Motta2015}, we excluded from our analysis features with a significance less than $3\sigma$. When more than one peak was present in the PDS, we identified the fundamental based on the QPO evolution along the outburst. We identified the peaks near half and twice the fundamental frequency as the subharmonic and second harmonic, respectively. We calculated the fractional amplitude (rms) of the QPO fundamental in six energy bands ($2-5.7$ keV, $5.7-7.7$ keV, $7.7-10.6$ keV, $10.6-15$ keV, $15-20.6$ keV and $20.6-44$ keV) to make an rms spectrum (rms as a function of energy).

We also produced a frequency-dependent phase lag spectrum (lag--frequency spectrum) between the $2-5.7$ keV and $5.7-15$ keV energy bands for each observation, following the method described in \citet{Vaughan1997} and \citet{Nowak1999}. To calculate the phase lags, $\Delta\phi$, at the QPO fundamental and (sub)harmonic, we averaged the phase lags over the width of each Fourier component, around its centroid frequency, $\nu_{0}\pm\rm{FWHM}/2$, where $\nu_{0}$ is the centroid frequency of the fundamental or the (sub)harmonic and FWHM is its full-width at half-maximum obtained from the Lorenzian fits. Throughout this paper, a positive phase lag means that the hard photons lag the soft photons. No correction for the dead-time-driven cross-talk effect \citep{Klis1987} was done because this effect was found to be negligible. The corresponding time lag, $\Delta\tau$, at a frequency $\nu$ is $\Delta\tau=\Delta\phi/2\pi\nu$.

In addition to the lag-frequency spectrum, we also calculated the energy-dependent phase lag for both the QPO fundamental and its harmonics (lag--energy spectrum), following the procedure described in \citet{Uttley2014}. The phase lags were calculated for the energy bands $4-5.7$ keV, $5.7-7.8$ keV, $7.8-10.6$ keV, $10.6-15$ keV, $15-20.6$ keV, $20.6-44$ keV, with reference to the softest band, $2-4$ keV.

\begin{figure*}
\gridline{\rotatefig{-90}{pds01.ps}{0.28\textwidth}{}
          \rotatefig{-90}{pds02.ps}{0.28\textwidth}{}
          \rotatefig{-90}{pds03.ps}{0.28\textwidth}{}
          }
\gridline{\rotatefig{-90}{pds04.ps}{0.28\textwidth}{}
          \rotatefig{-90}{pds05.ps}{0.28\textwidth}{}
          \rotatefig{-90}{pds06.ps}{0.28\textwidth}{}
          }
\caption{Examples of PDS and phase lag spectra as a function of Fourier frequency for six observations of GX 339--4 with different QPO frequencies. The power spectra were calculated in the full energy band and fitted with a function consisting of the sum of several Lorentzians. The phase lag spectra were calculated for the $5.7-15$ keV band relative to the $2-5.7$ keV band. Positive lags mean that hard X-ray photons ($5.7-15$ keV) lag the soft ones ($2-5.7$ keV). Observation ID and QPO fundamental frequency are shown for each panel in the top left corner. The vertical solid and dashed lines indicate the ranges over which the QPO fundamental and second-harmonic lags are averaged ($\nu_{0}\pm\rm{FWHM}/2$), respectively.}
\label{fig:pds}
\end{figure*}

\begin{figure}
\includegraphics[width=0.45\textwidth]{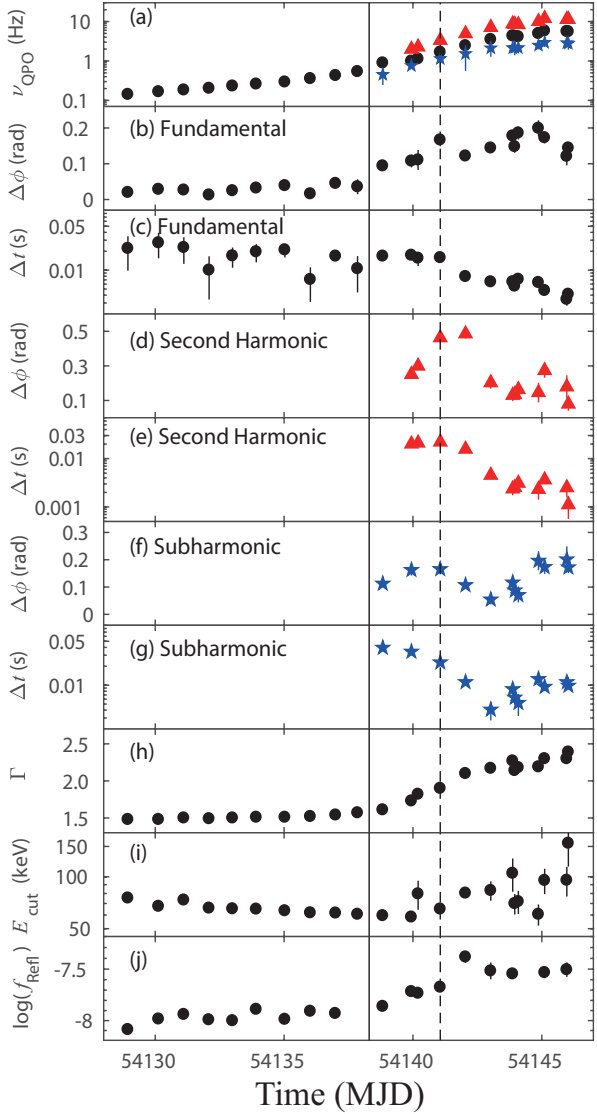}
\caption{Evolution of the QPO frequencies, phase and time lags and the main spectral parameters of GX 339--4 as a function of time. (a) The centroid frequency of the QPO fundamental (circles). If present, second harmonic (red triangles) and subharmonic (blue stars) frequencies are also plotted. (b)-(g) Phase and time lags, $\Delta\phi$ and $\Delta t$, respectively, at the QPO fundamental, second harmonic and subharmonic. The next two panels show parameters of the spectral fit results of \citet{Motta2009}. (h) Power-law photon index, $\Gamma$. (i) High-energy cut-off, $E_{\rm cut}$, in keV. Note that-in Figure 6 of \citet{Motta2009}-the parameter values of the high-energy cut-off were averaged across some observations with similar hardness. Here we did not average them. (j) Reflection flux in units of $\rm erg~cm^{-2}~s^{-1}$. The values are taken from \citet{Plant2014}. The solid line marks the transition from the LHS to the HIMS. The dotted line marks the observation corresponding to the break in the lag--frequency and lag--$\Gamma$ relations (see Figure \ref{fig:phase_fund} and corresponding text below).}
\label{fig:time_elv}
\end{figure}

\section{Results} \label{sec:results}

\begin{figure*}[!ht]
\plottwo{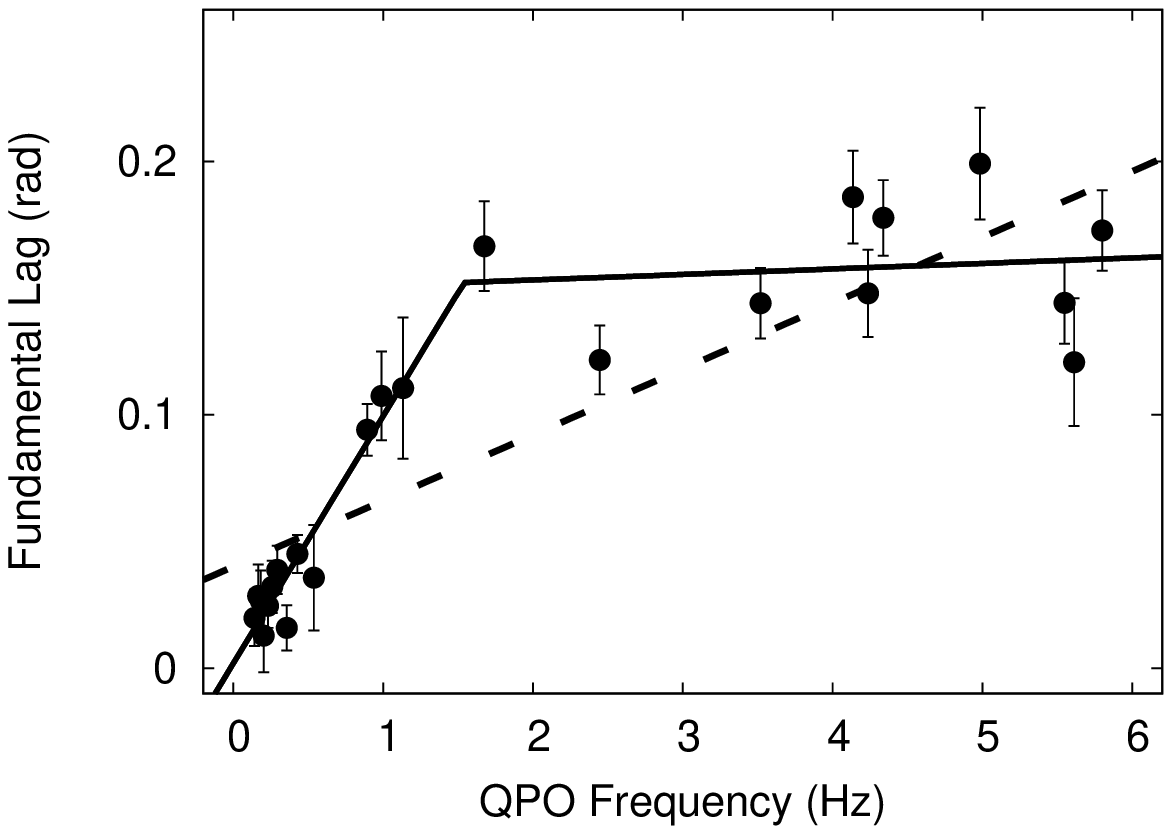}{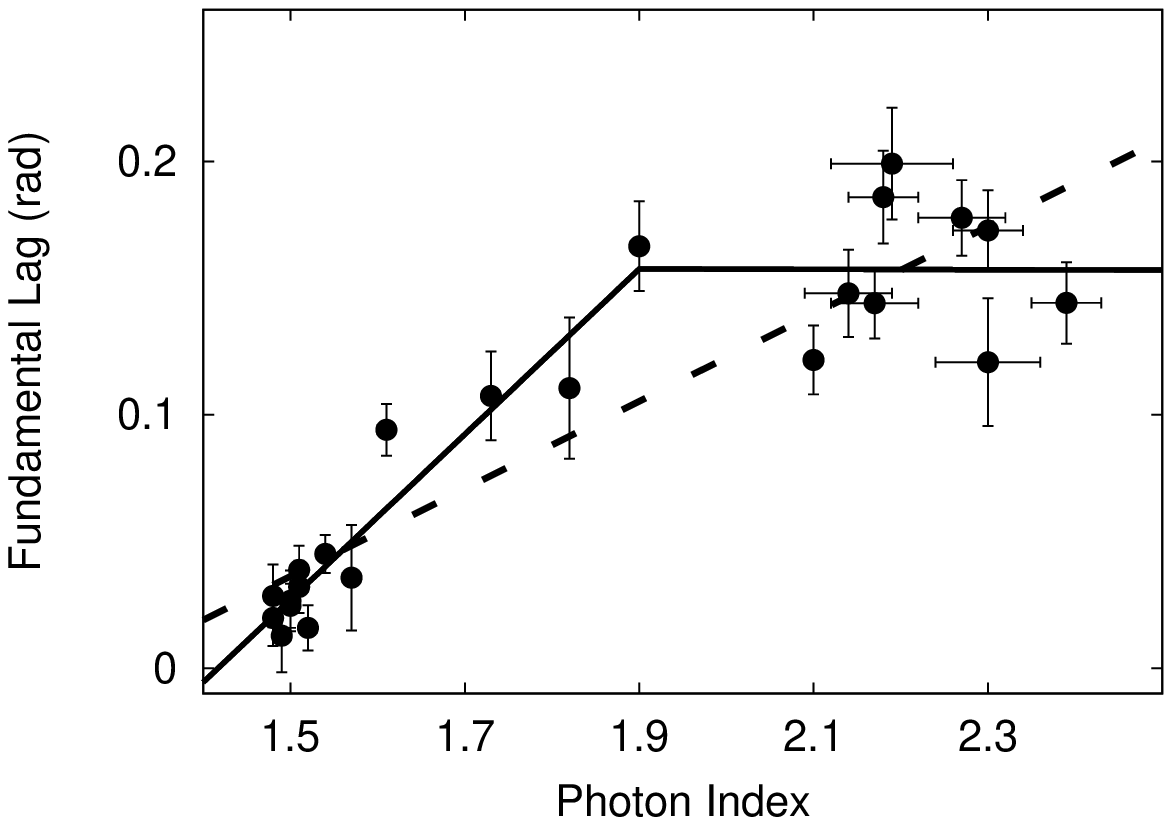}
\caption{Phase lags at the QPO fundamental vs. QPO frequency (\emph{left panel}) and photon index (\emph{right panel}) for GX 339--4. Each point represents a single \emph{RXTE} observation. In both panels, the solid line represents the best fit to the points with a broken line, while the dashed line represents the best fit with a straight line (see text for details).}
\label{fig:phase_fund}
\end{figure*}

\begin{figure*}
\includegraphics[width=\textwidth]{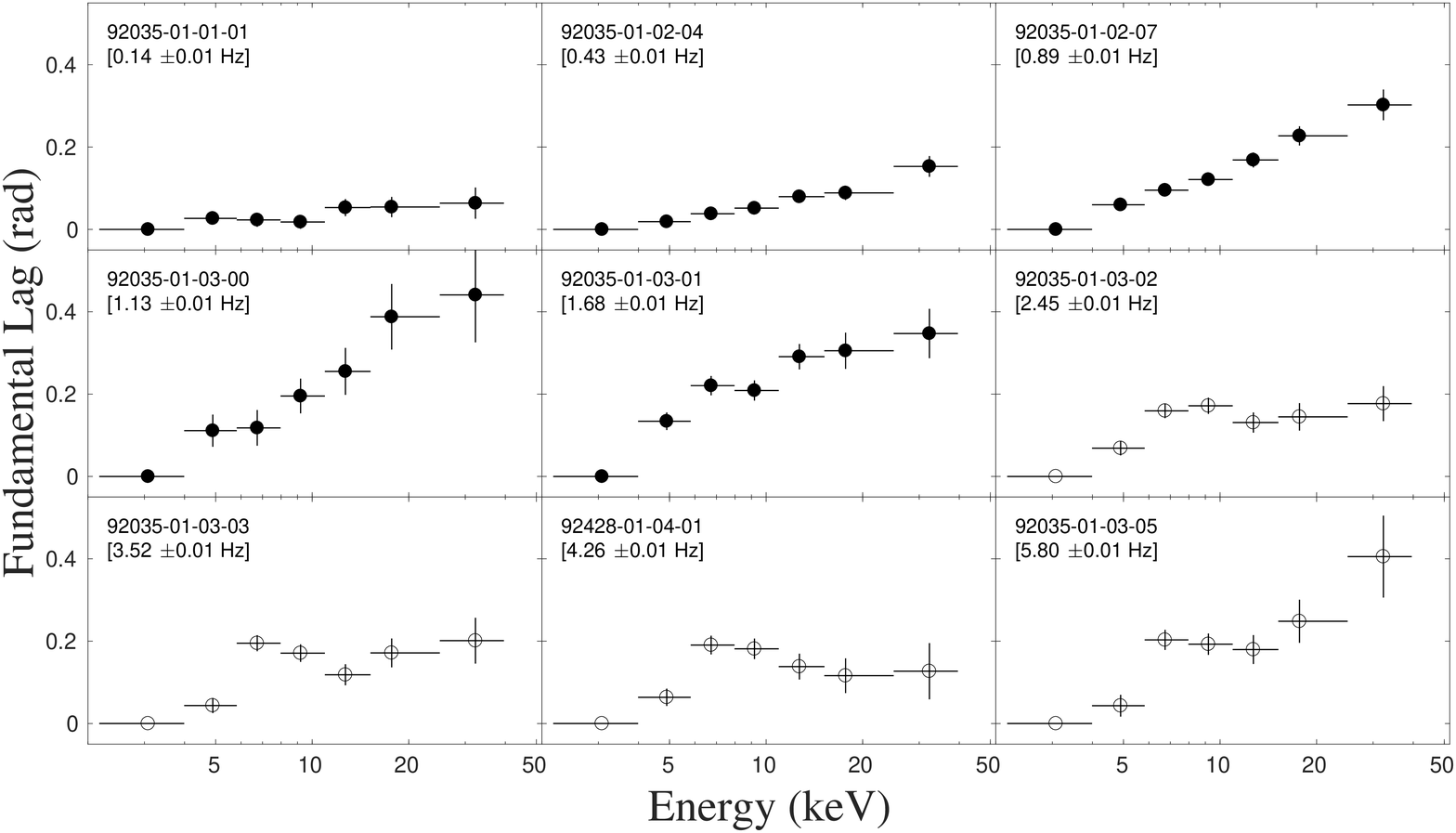}
\caption{Energy dependence of the phase lags at the QPO fundamental for observations of GX 339--4 with different QPO frequencies. Observation ID and QPO frequency are indicated in each panel. The observations with QPO frequencies below and above $\sim1.7$ Hz are shown with different markers.}
\label{fig:phase_energy}
\end{figure*}

In Figure \ref{fig:pds}, we show the PDS (upper panels) and lag--frequency spectra (lower panels) for six representative observations of GX 339--4 with different QPO frequencies. In most cases, we observe a dip-like feature near the QPO fundamental frequency in the lag--frequency spectra; a similar feature was also reported by \citet{Pahari2013} in a timing analysis of GRS 1915+105. This feature seems to be more obvious at high QPO frequency and is not always centered on, but sometimes deviates slightly from, the centroid frequency of the fundamental. In addition, we also find that, if a second harmonic is present, a peak-like feature is visible as well in the lag--frequency spectra around the second harmonic frequency.

Figure \ref{fig:time_elv} shows the evolution of the QPO frequency, phase and time lags, together with the main spectral parameters, as a function of time. The values of the power-low photon index and the high-energy cutoff are taken from \citet{Motta2009} (see Table 3 in the electronic version of the article), who fitted the continuum with a multi-colour disc blackbody and a cut-off power law, and the reflection component with a gaussian line and a smeared edge. The values of the reflection flux are from \citet{Plant2014} \footnote{Note that \citet{Plant2014} did not analyse all the observations we used in our paper}, who used the same continuum model as \citet{Motta2009}, but used a relativistically broadened reflection component instead of the gaussian line and the smeared edge. Although the reflection models in these two papers are different, the evolution of the high-energy cut off and power-law index are consistent with each other.

From MJD 54129 to 54138, the power-law photon index increases slightly from $\sim1.5$ to $\sim1.6$, while the high-energy cut-off decreases monotonically from $\sim75$ to $\sim60$ keV, which is typical for the LHS. During the LHS, the reflection flux shows a slightly increasing trend. The phase lags at the QPO fundamental are low, and no significant harmonic component of this QPO is found in the PDS. Starting from MJD 54138, corresponding to the observation with a QPO frequency of $\sim0.54$ Hz, the source undergoes a transition to the HIMS with a fast increase in the power-law photon index and the reflection flux. The high-energy cut-off stops decreasing and starts increasing. The phase lags at the fundamental increase rapidly from $\sim0.05$ to $\sim0.17$ rad in a few days. During the transition, a second harmonic and subharmonic of the QPO begin to appear in the PDS, with their phase lags increasing with time. After MJD 54142, corresponding to the observation with a QPO frequency of $\sim1.7$ Hz, the power-law photon index slowly increases but the reflection flux remains more or less constant. The QPO fundamental phase lags do not change a lot. However, both the subharmonic and the second-harmonic phase lags drop quickly to a local minimum, and then the subharmonic phase lags increase again, while the second-harmonic phase lags show no clear trend.

Before MJD 54142, the time lags at the QPO fundamental remain more or less constant, while after MJD 54142, the time lags decrease as the QPO frequency increases. Figure \ref{fig:time_elv} shows that both for the sub- and second harmonic of the QPO, the time evolution of the phase and time lags is similar. Because, for the purpose of this paper, both the phase and time lags of the fundamental and harmonic components of the QPO provide the same information, in the rest of the paper we only show results of the phase lags.

In Figure \ref{fig:phase_fund}, we show the phase lags at the QPO fundamental as a function of QPO frequency (\emph{left}) and power-law photon index (\emph{right}), respectively. Since the QPO frequency increases monotonically with power-law photon index, the evolution of the phase lags with QPO frequency and power-law photon index are quite similar. A clear break is present in both the lag--frequency and lag--$\Gamma$ relations. At QPO frequency less than $\sim1.7$ Hz ($\Gamma\sim1.9$), the phase lags generally increase with QPO frequency (photon index), but appear to remain more or less constant at higher frequency. To check whether there is a significant change in the phase lag at a QPO frequency of $\sim1.7$ Hz, we fitted the two relations both with a broken line and a straight line. We compared the fits using an F-test, and find that the broken line fit is better than the straight line fit at a confidence level of $5~\sigma$ and $3~\sigma$ for the lag--frequency and lag--$\Gamma$ relations, respectively, suggesting that the behavior of the fundamental phase lag changes significantly. The break obtained from the fit is at $\nu_{\rm{QPO}}=1.55\pm0.15$ Hz for the lag--frequency relation and $\Gamma=1.90\pm0.10$ for the lag--$\Gamma$ relation. Even if the best-fitting value of the break is at a frequency of 1.55 Hz, we will use 1.7 Hz in the rest of the paper because that is the value of the QPO frequency that is closest to the break.

\begin{figure}
\includegraphics[width=0.5\textwidth]{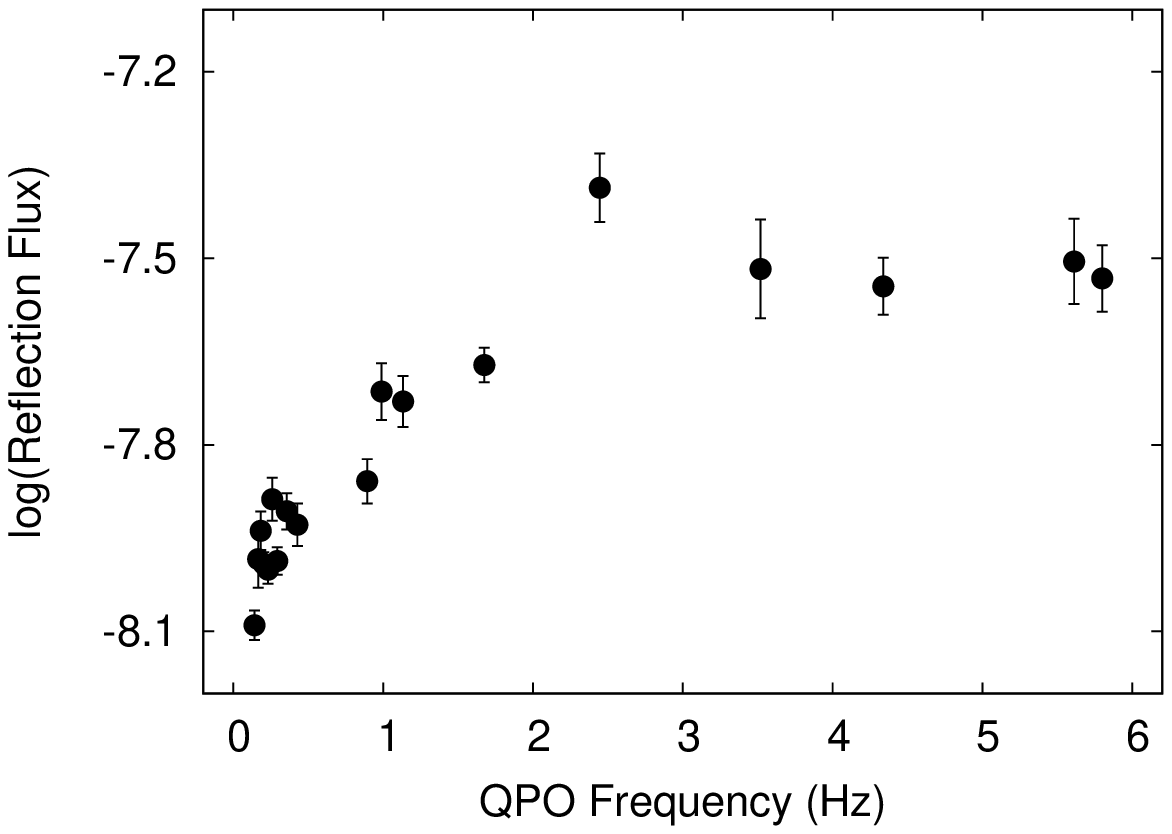}
\caption{Reflection flux vs. QPO frequency. The data of the reflection flux are from \citet{Plant2014}. The y-axis is in units of $\rm erg~cm^{-2}~s^{-1}$.}
\label{fig:fre_rflux}
\end{figure}

To better understand the differences in phase lag behavior between QPOs below and above $\sim1.7$ Hz, we measured the phase lags at the QPO fundamental as a function of energy for each observation. In Figure \ref{fig:phase_energy}, we show representative energy-dependent phase lag for nine observations with different QPO frequencies. We find that for the QPOs with centroid frequency below $\sim1.7$ Hz, the phase lags always increase monotonically with energy. We fitted the lag--energy spectra with a log-linear function of $\Delta\phi(E)=\alpha {\rm log} (E)+\beta$. We find that the slope, $\alpha$, increases from $0.02\pm0.02$ for $\nu_{\rm{QPO}}=0.17\pm0.01$ Hz to $0.43\pm0.07$ for $\nu_{\rm{QPO}}=1.13\pm0.01$ Hz. On the contrary, for the QPOs with centroid frequency above 1.7 Hz, the phase lags show a non-monotonic behavior with energy: the phase lags first increase with energy until about $\sim6.5$ keV and then decrease, while at energies above $\sim12$ keV the phase lags remain more or less constant or increase again. When the QPO frequency is higher than $\sim1.7$ Hz, a relatively broad feature is always observed in the energy-dependent phase lag at around 6.5 keV, where the iron line is present in the energy spectrum \citep{Miller2007}.

Such reflected feature in the lag-energy spectra of the QPO when its frequency is above $\sim1.7$ Hz suggests that the phase lags may be related to the reflected emission \citep[see, e.g.,][]{Kotov2001}. If this is the case, the constant phase lags when the QPO frequency is higher than $\sim1.7$ Hz foretell that the reflection component in these observations should be constant. In Figure \ref{fig:fre_rflux} we show the reflection flux as a function of QPO frequency. Indeed, the reflection flux increases with QPO frequency first, and then remains approximately constant above the break. The evolution of the reflection flux with QPO frequency is quite similar to that of the QPO phase lags, with a clear break near 1.7 Hz.

\begin{figure*}
\includegraphics[width=\textwidth]{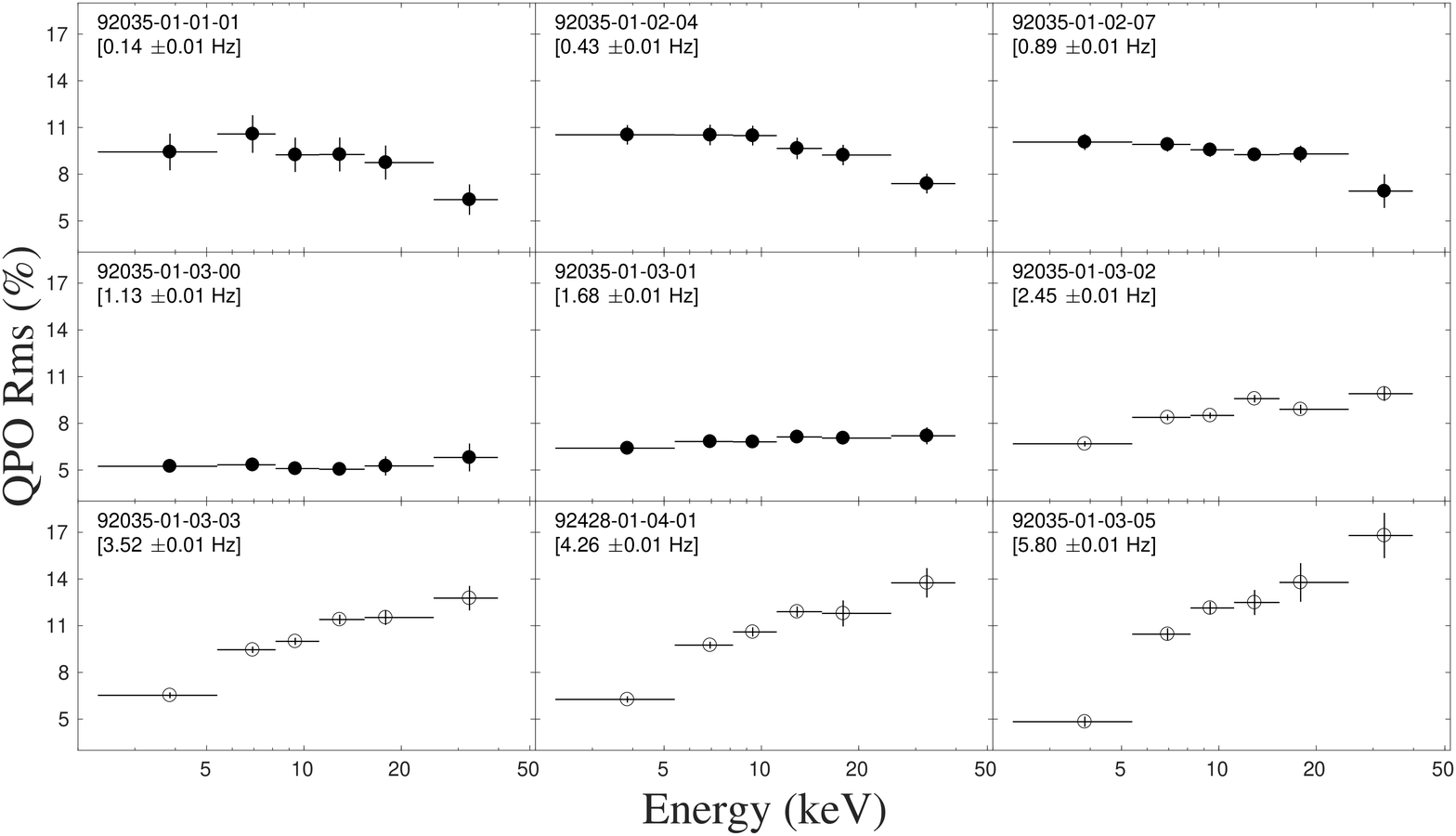}
\caption{Energy dependence of the QPO rms for observations of GX 339--4 with different QPO frequencies. Observation ID and QPO frequency are indicated in each panel. The observations with QPO frequencies below and above $\sim1.7$ Hz are shown with different markers.}
\label{fig:rms_energy}
\end{figure*}

\begin{figure}
\includegraphics[width=0.5\textwidth]{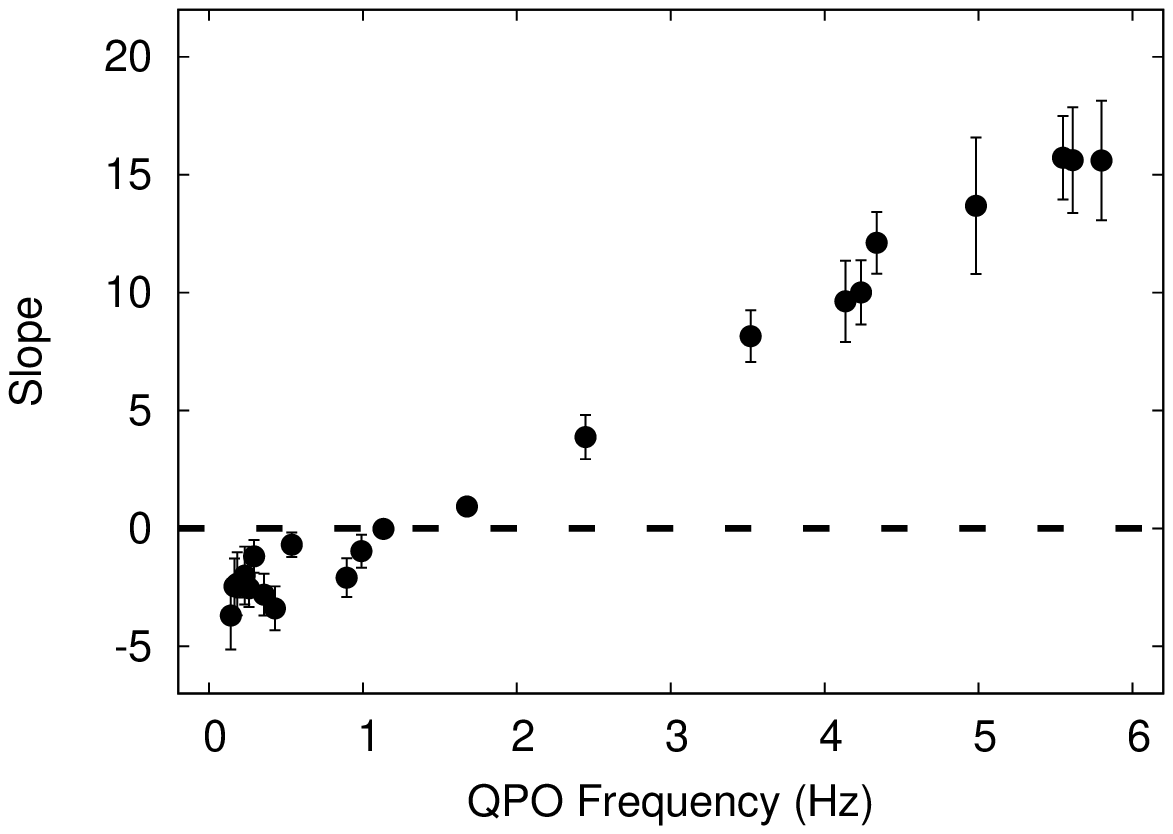}
\caption{The slope of the energy dependence of the QPO rms as a function of QPO frequency, fitted with the same log-linear function that we used for the lag--energy spectra of the QPO fundamental (see text for details). The slope increases monotonically with QPO frequency.}
\label{fig:rms_fit}
\end{figure}

We also find differences between QPOs below and above $\sim1.7$ Hz in the energy dependence of the QPO fractional rms, as shown in Figure \ref{fig:rms_energy}. When the centroid frequency of the QPO is below $\sim1.7$ Hz, the QPO rms decreases with energy, when the QPO frequency is close to $\sim1.7$ Hz, the rms spectrum of the QPO is almost flat, while when the centroid frequency of the QPO is above $\sim1.7$ Hz, the QPO rms increases with energy. We fitted the rms spectra with the same log-linear function that we used for the lag--energy spectra of the QPO fundamental. Figure \ref{fig:rms_fit} shows the slope of the rms spectra as a function of QPO frequency. We find that the slope increases from negative to positive values ($-3.7\pm1.4$ to $15.7\pm1.8$) as the increasing QPO frequency. All the results above indicate that some changes occur in the system at a QPO frequency of $\sim1.7$ Hz. It is worth noting that the QPO frequency increases very quickly from $\sim1.7$ Hz to $\sim2.4$ Hz, within 1 day. Because of this, we cannot assert accurately the QPO frequency where these changes occur.

\begin{figure*}
\plottwo{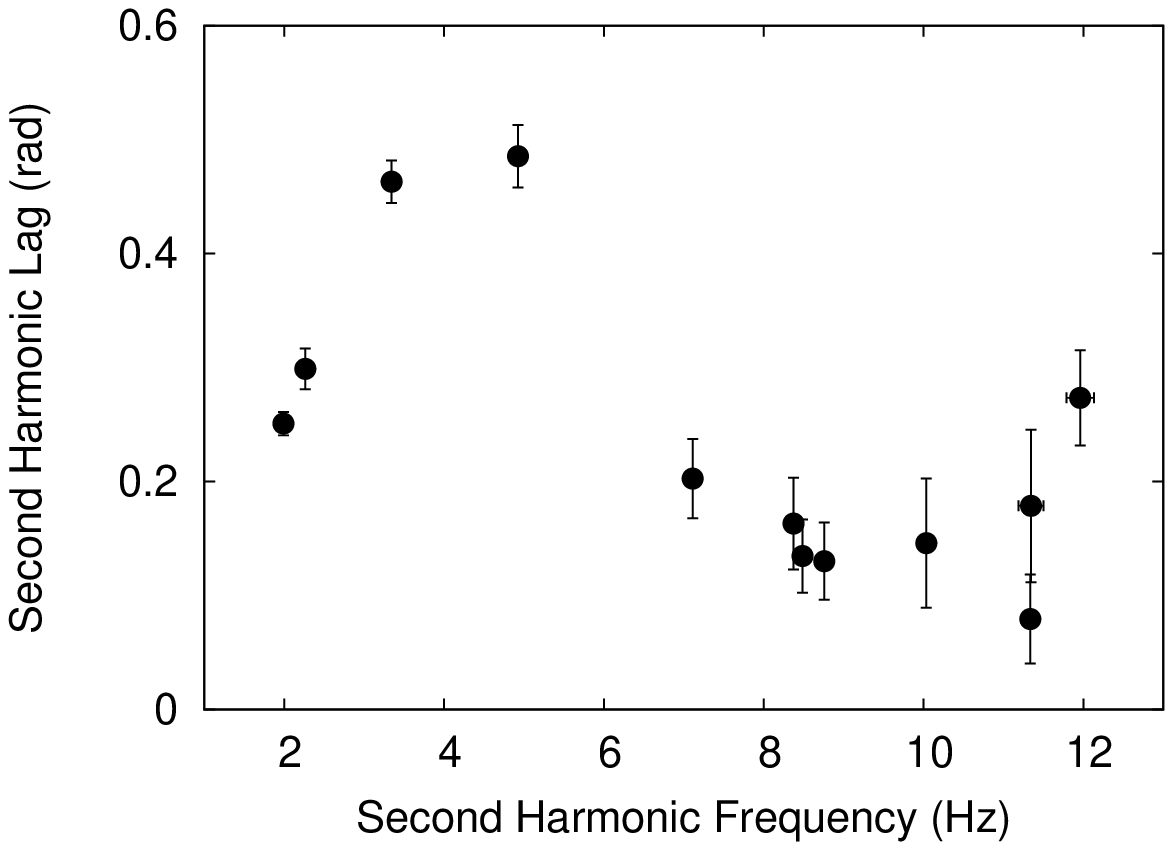}{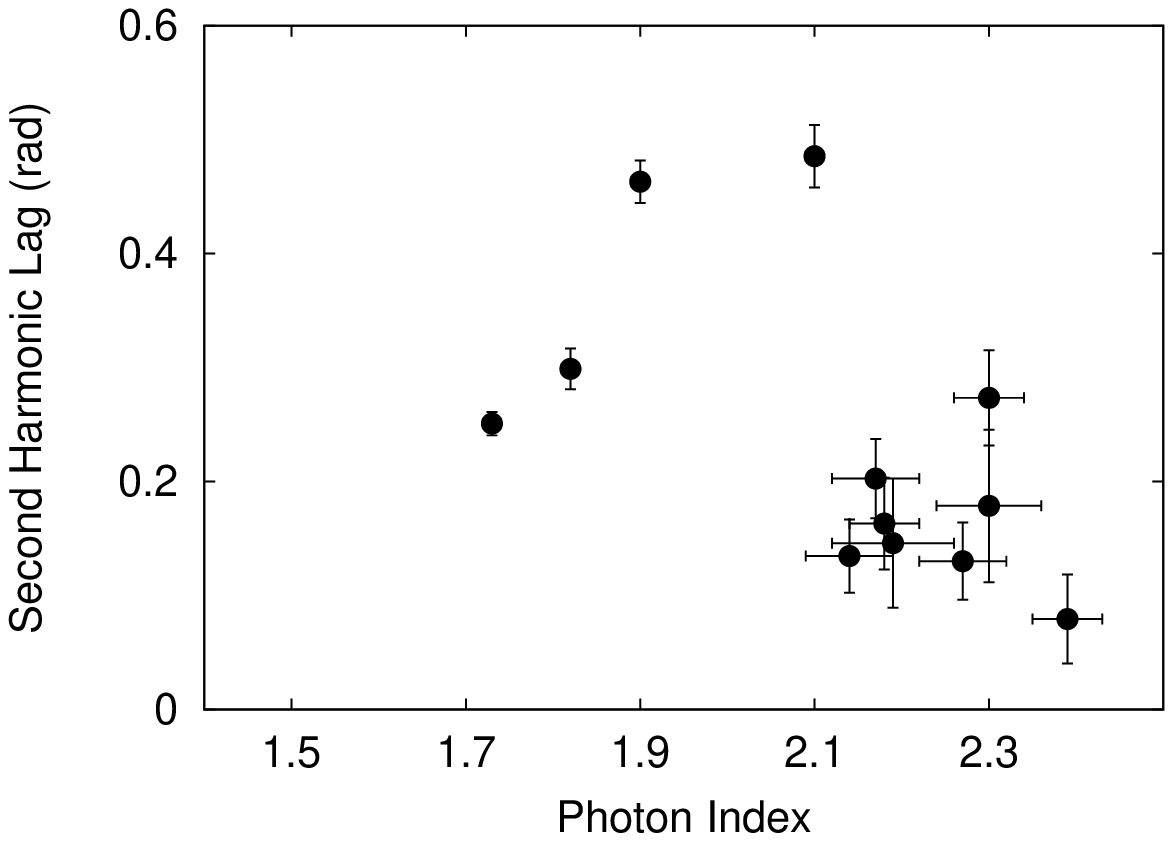}
\caption{Phase lags at the second-harmonic frequency vs. second-harmonic frequency (\emph{left panel}) and photon index (\emph{right panel}) for GX 339--4. Each point represents a single \emph{RXTE} observation.}
\label{fig:phase_second}
\end{figure*}

As shown in Figure \ref{fig:pds}, the second harmonic and subharmonic appear during the transition from the LHS to the HIMS. Figure \ref{fig:phase_second} shows the phase lags at the second harmonic as a function of second-harmonic frequency (\emph{left}) and power-law photon index (\emph{right}), respectively. The phase lags show a local maximum at a second-harmonic frequency of $\sim4.9$ Hz, and then drop as the frequency increases further. At high second-harmonic frequency, the phase lags appear to increase again, but this increase is not apparent in the lag-$\Gamma$ relation.

\begin{figure*}
\includegraphics[width=\textwidth]{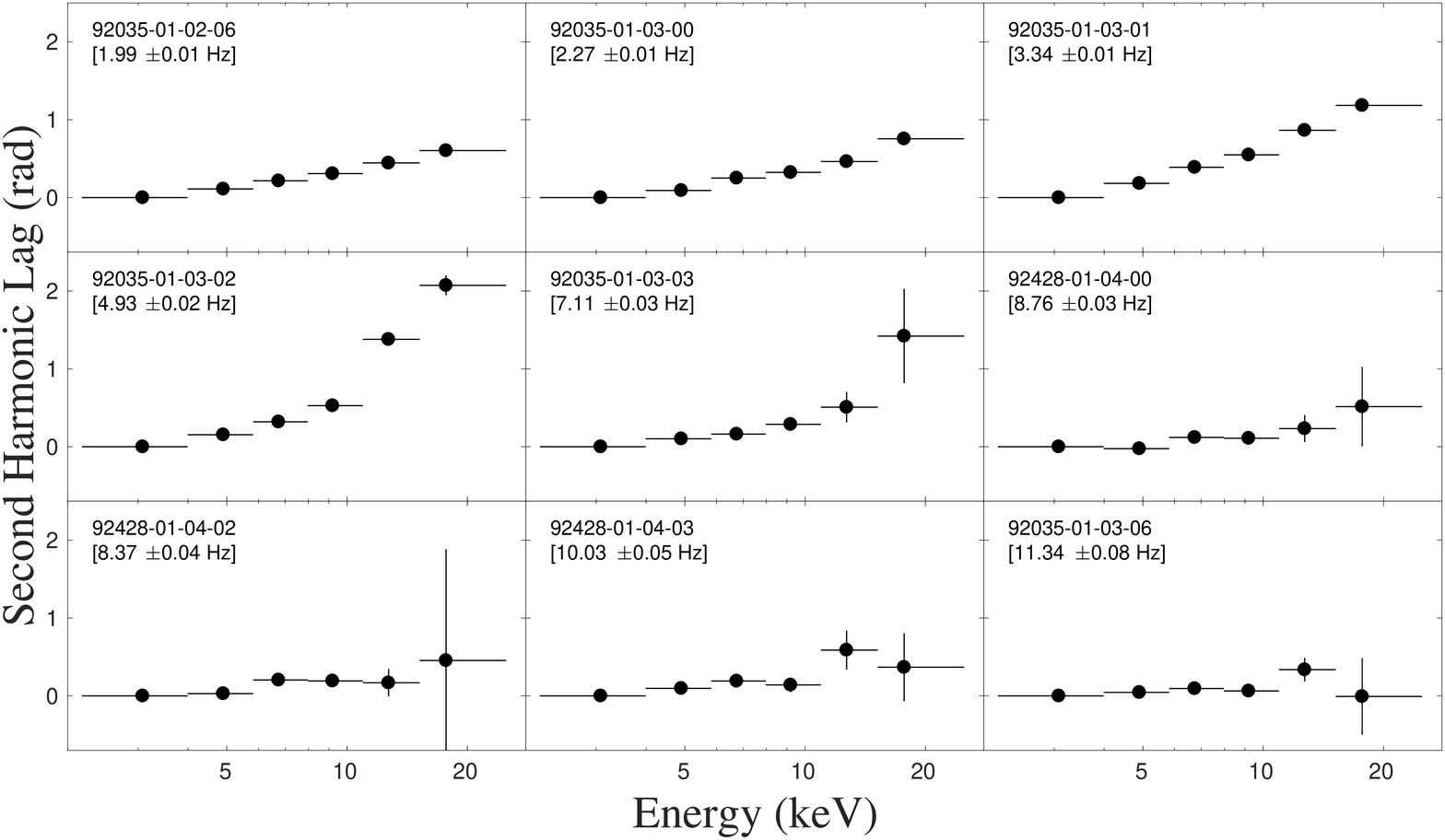}
\caption{Energy dependence of the phase lags at the second harmonic for GX 339--4. Observation ID and second-harmonic frequency are indicated in each panel.}
\label{fig:phase_energy_second}
\end{figure*}

\begin{figure}
\includegraphics[width=0.5\textwidth]{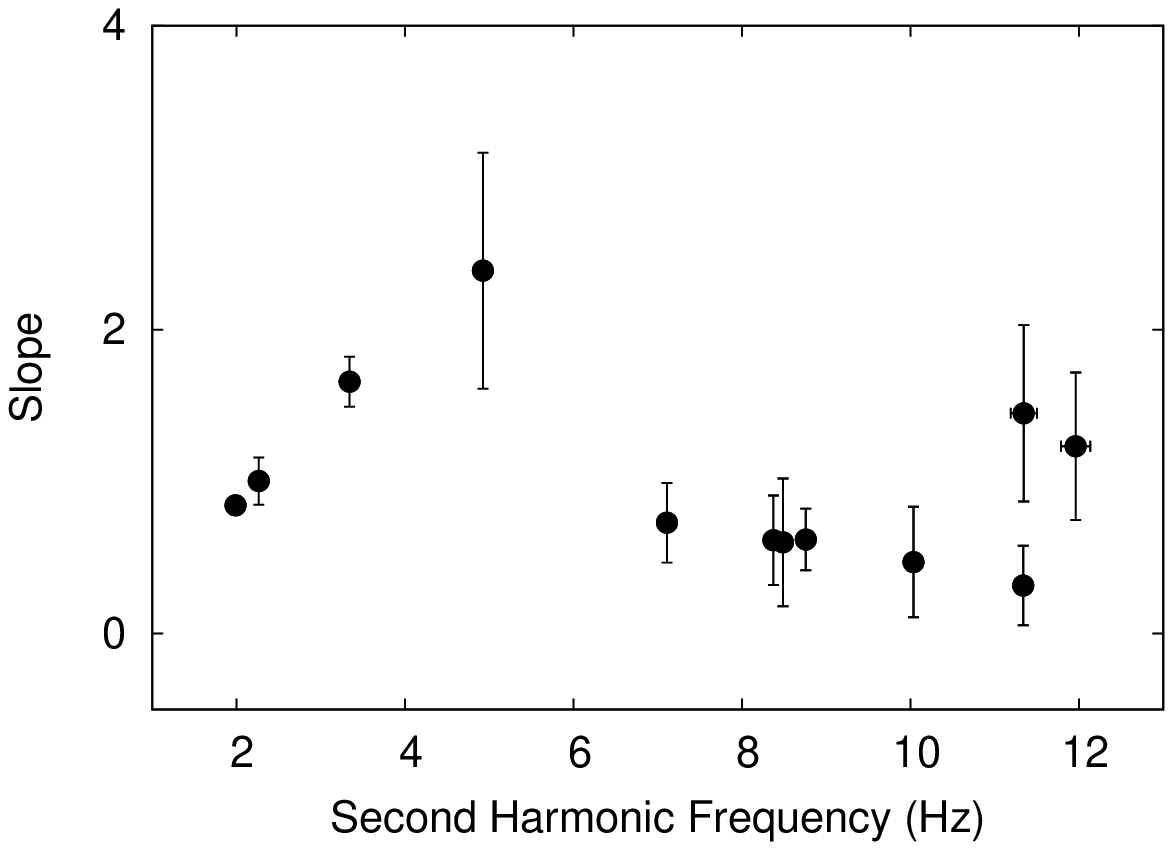}
\caption{The slope of the energy dependence of the phase lags at the second harmonic as a function of second-harmonic frequency, fitted with the same log-linear function that we used for the lag--energy spectra of the QPO fundamental (see text for details). There is a clear change in the slope at a second-harmonic frequency of $\sim6$ Hz.}
\label{fig:phase_fit_second}
\end{figure}

Figure \ref{fig:phase_energy_second} shows the energy-dependent phase lag spectra for the QPO second harmonic. Due to the small amplitudes of the (sub)harmonic compared to the fundamental, and the low signal-to-noise ratio, at high energies the phase lags of the (sub)harmonic show large uncertainties. Therefore, in this case we ignore the data above 20 keV. We find that the phase lags at the second harmonic increase with energy at second-harmonic frequencies less than $\sim6$ Hz, while it becomes flat at second-harmonic frequencies larger than $\sim6$ Hz. The slope of the lag-energy spectra of the second harmonic as a function of second-harmonic frequency is shown in Figure \ref{fig:phase_fit_second}. The frequency at which the second harmonic phase-lag behavior changes is not 2 times that of the QPO fundamental frequency.

\begin{figure*}
\plottwo{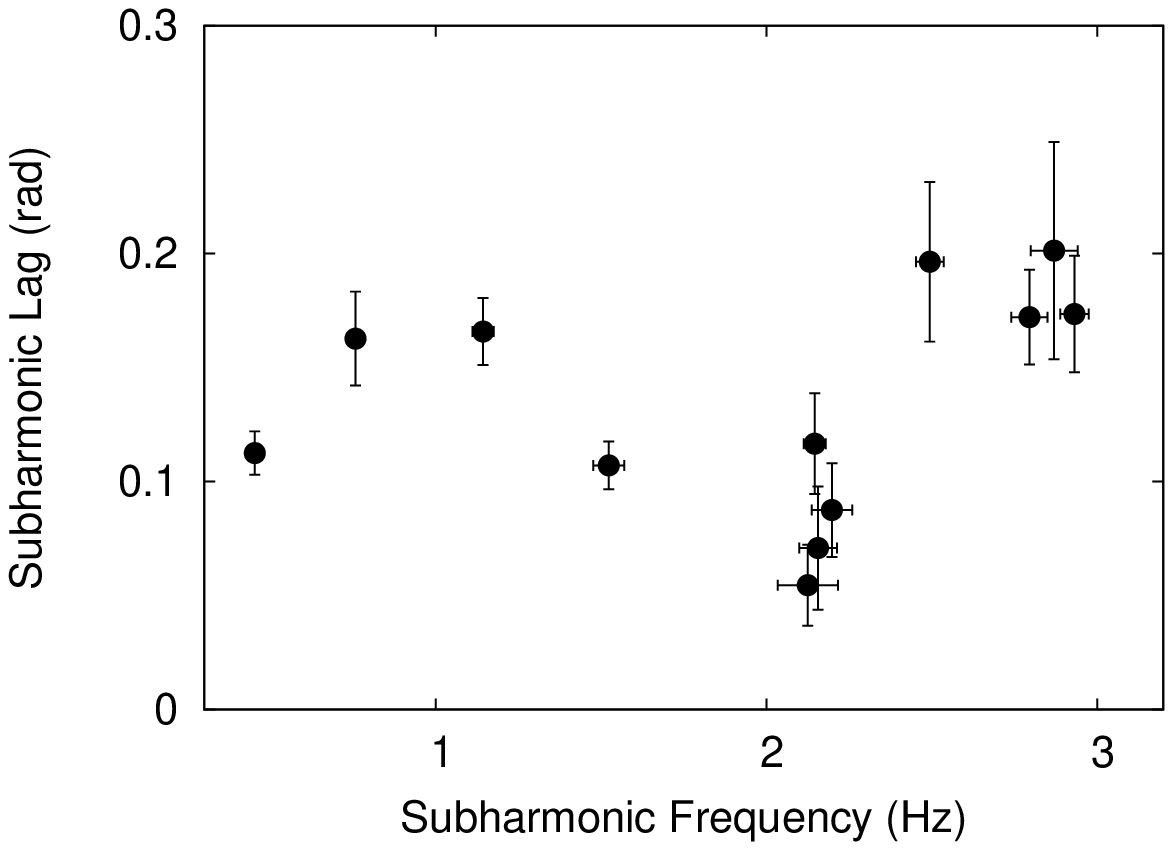}{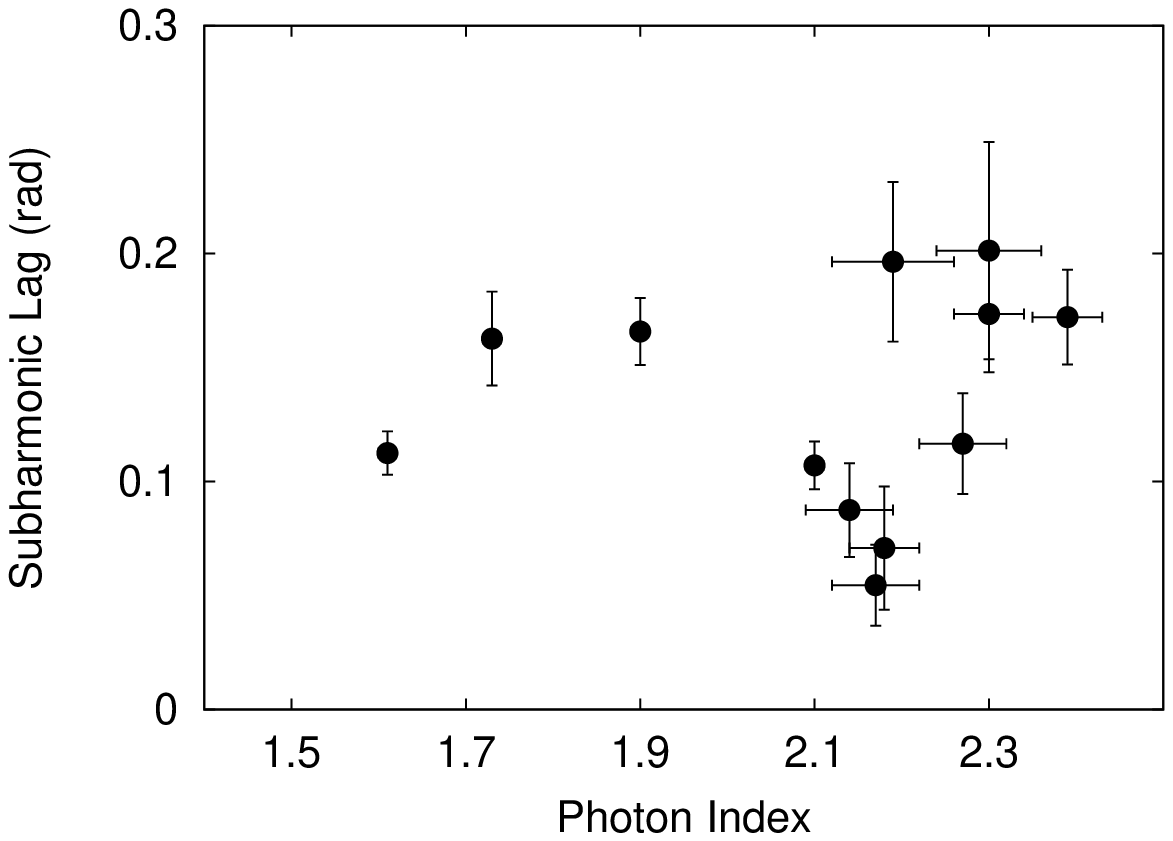}
\caption{Phase lags at the subharmonic frequency vs. subharmonic frequency (\emph{left panel}) and photon index (\emph{right panel}) for GX 339--4. Each point represents a single \emph{RXTE} observation.}
\label{fig:phase_sub}
\end{figure*}

\begin{figure*}
\includegraphics[width=\textwidth]{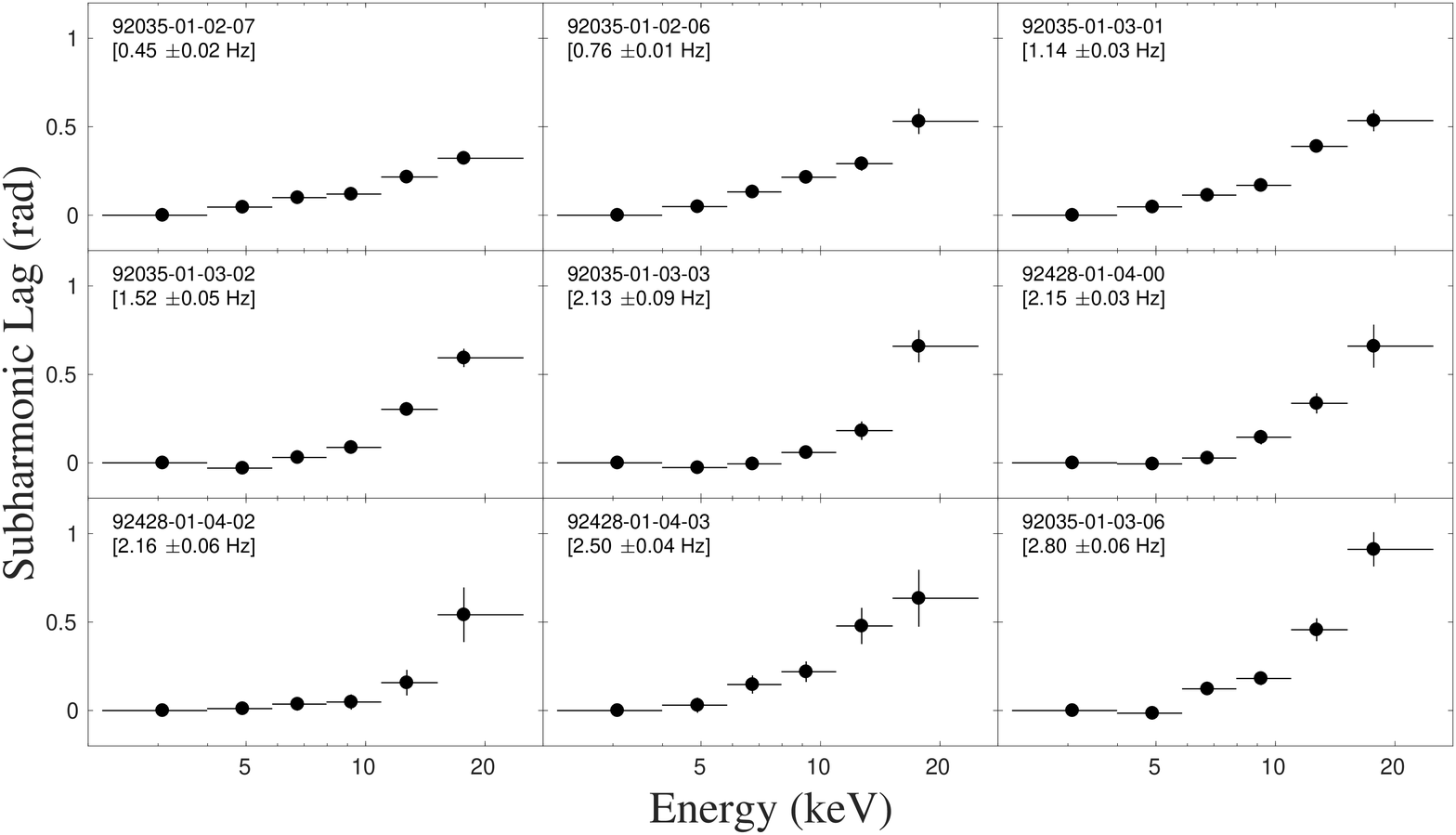}
\caption{Energy dependence of the phase lags at the subharmonic for GX 339--4. Observation ID and subharmonic frequency are indicated in each panel.}
\label{fig:phase_energy_sub}
\end{figure*}

Figure \ref{fig:phase_sub} shows the phase lags at the subharmonic as a function of the subharmonic frequency (\emph{left}) and power-law photon index (\emph{right}), respectively, and Figure \ref{fig:phase_energy_sub} shows the representative energy-dependent phase lag spectra at the subharmonic. Figure \ref{fig:phase_second} and \ref{fig:phase_sub} show that the phase lags of the subharmonic and the second harmonic of the QPO behave in a similar way. On the contrary, the lag--energy spectra of the subharmonic and the second harmonic are quite different (Figure \ref{fig:phase_energy_second} and \ref{fig:phase_energy_sub}): the phase lags of the subharmonic do not turn to flat with energy at high subharmonic frequencies but, instead, always increase with energy.

\section{Discussion} \label{sec:discussion}

We present the evolution of the phase lags associated with the type-C QPO in GX 339--4 during the rising phase of its 2006/2007 outburst. We find, for the first time in this source, the phase lags at the type-C QPO fundamental show very different behavior when the QPO frequency is below or above $\sim1.7$ Hz, both as a function of QPO frequency and energy (see Figure \ref{fig:phase_fund} and \ref{fig:phase_energy}). Interestingly, we also find a change in the shape of the QPO rms spectra at a QPO frequency of $\sim1.7$ Hz (Figure \ref{fig:rms_energy}). During the transition from the LHS to the HIMS, which takes place at a QPO frequency of $\sim0.54$ Hz, the second harmonic and subharmonic appear in the PDS. We observe a clear change in the shape of the energy-dependent phase lag at the second harmonic, where the phase lags increase with energy at low second-harmonic frequency and become flat with energy at high second-harmonic frequency (Figure \ref{fig:phase_energy_second}). No significant change occurs in the shape of the energy-dependent phase lag at the subharmonic (Figure \ref{fig:phase_energy_sub}).

Recently, \citet{Eijnden2017} found that the phase lags at the type-C QPO fundamental depend significantly on source inclination: at low QPO frequencies, all sources show slightly hard lags; at high frequencies, low-inclination sources still display hard lags while high-inclination sources turn to soft lags. As shown in Figure \ref{fig:time_elv}, we find that the type-C QPOs in GX 339--4 only display hard lags, which is consistent with the fact that GX 339--4 is a low-inclination system.

Several models have been proposed to explain the physical origin of the lags observed in BHTs. The hard lags could originate from Compton up-scattering of soft seed photons by hot electrons in a uniform corona, since the high-energy photons have to undergo more scatterings to gain energy \citep[e.g.,][]{Miyamoto1988,Cui1999,Nowak1999}. In this scenario, the lags would be proportional to the light-crossing time of the corona \citep{Nowak1999}. Alternatively, the lags could be due to the Comptonization of soft disc photons in the jet \citep{Kylafis2008}. The reflection process could also give rise to hard lags \citep{Kotov2001}. The delayed photons travel from the direct emission region to the reflector first, and then to the observer. In this case, the lags are of the order of the light-crossing time between the direct emission region and the reflector. More importantly, in this scenario, the lag--energy spectra should contain a broad iron line feature with an equivalent width much larger than that in the averaged spectrum \citep{Kotov2001}. Note that, both the Comptonization and reflection models explain the hard lags as light travel time effect and could give rise to time lags of the order of a few milliseconds. Longer time lags could be due to the inward propagation of mass accretion fluctuations. These fluctuations are introduced over a wide range of radii and propagate inward through the disc \citep{Lyubarskii1997,Kotov2001,Arevalo2006}. The propagation of accretion rate fluctuations is on a time-scale of seconds, much longer than the expected light-crossing time \citep{Uttley2011}. However, the maximum phase lags in GX 339--4 is $\sim0.2$ rad. For a 4 Hz type-C QPO, this corresponds to a time lag of $\sim8$ ms, which is incompatible with the propagation time-scale. Moreover, recent results show that the type-C QPO most likely originates from the inner flow (corona), rather than the disc \citep[see][]{Belloni2011}. Assuming that the phase lags at the QPO frequency are dominated by the QPO itself, the QPO phase lags should also originate from a process related to the corona. However, in the propagating fluctuation scenario, the fluctuations are produced only in the accretion disc and propagate inward before they reach the corona \citep{Lyubarskii1997}. It is therefore unlikely that this mechanism is responsible for the observed phase lags.

An important result of our analysis is the change in the shape of the lag--energy spectra at a QPO frequency of $\sim1.7$ Hz. At low QPO frequencies ($<1.7$ Hz), the phase lags smoothly increase with energy. Such a logarithmic lag--energy dependence can be naturally explained both by the Comptonization and propagating fluctuation models. In the former, the higher energy photons undergo more scatterings, and thus have a longer path length to travel before leaving the corona. In the latter, the emission coming from the inner disc is harder than that of the outer regions. Therefore, the high-energy photons have a longer propagation time from the outer to the inner disc than the low-energy photons. At high QPO frequencies ($>1.7$ Hz), a prominent feature at the energy of the Fe K$\alpha$ line ($\sim6.5$ keV) is present in the lag--energy spectra, and above $\sim12$ keV, the energy of the reflection hump, the phase lags appear to increase again. The appearance of these reflected features when the reflection flux is high provides strong evidence that, at least in these cases, the phase lags are associated with the reflection component.

Another important result is the break observed in Figure \ref{fig:phase_fund}: the QPO phase lags first increase as the QPO frequency increases, and then remain more or less constant when the QPO frequency is higher than $\sim1.7$ Hz. A similar trend is seen in Figure \ref{fig:fre_rflux}, between the reflection flux and the QPO frequency. We note that the reflected feature seen in the lag--energy spectra when the QPO frequency is above $\sim1.7$ Hz also has a near constant phase lag of $\sim0.2$ rad (see Figure \ref{fig:phase_energy}). All these support the idea that when the QPO frequency is above $\sim1.7$ Hz, the QPO phase lags are dominated by the reflection process. The break in Figure \ref{fig:phase_fund} and \ref{fig:fre_rflux} happens at MJD 54142. We note that after this time, the phase lags of the broad-band component ($0.008-5$ Hz) also remain approximately constant (see \citealt{Altamirano2015}, Figure 2B). The completely different phase-lag behaviour between QPOs below and above $\sim1.7$ Hz suggests that the QPO phase lags originate from different mechanisms, with the break indicating the moment of the change in the dominant mechanism of the phase lags.

The increase of the reflection flux with QPO frequency below the break in Figure \ref{fig:fre_rflux} is consistent with a scenario in which the disc moves inward as the source luminosity increases: based on the truncated disc model \citep{Esin1997}, the inner edge of the disc is truncated at some radius larger than the inner-most stable circular orbit (ISCO) in the LHS. As the source luminosity increases, the disc gradually moves inward towards the black hole, leading to an increase in the disc component and QPO frequency \citep{Ingram2009}. As the inner disc radius decreases, the solid angle it subtends below the corona increases, resulting in more hard photons irradiating the disc, and hence a stronger reflection component \citep{Done2007}. When the reflection component reaches a critical value, the QPO phase lags would be dominated by the reflection process, and a reflected feature would appear in the lag--energy spectra. The lack of any obvious reflection-related feature in the lag--energy spectra of the QPO when its frequency is below $\sim1.7$ Hz could be due to the fact that the reflection component at low luminosity is weak, and the QPO lags here may be dominated by another mechanism.

The saturation of the QPO phase lags with QPO frequency coincides with the saturation of the reflection flux with QPO frequency, implying that the region giving rise both to the lags and the reflection component reaches a stable configuration. One possibility is the disc approaching the ISCO near the break. In this case, the solid angle of the disc subtended by the corona remains constant, assuming that the coronal height does not change. The main objection to this scenario is that the QPO frequency still increases above the break. In the Lense-Thirring procession model, the increasing QPO frequency implies that the disc continues moving inwards \citep{Ingram2009}. In addition, we note that the time lags of the QPO gradually decrease after the break. Assuming, as we discussed earlier, that the lags are indeed dominated by the reflection process, this implies that the path from the primary source to the reflector decreases. Alternatively, assuming the disc still moves inward after the break, the constant reflection flux could be due to a change in the coronal height. An increase in coronal height decreases the portion of Comptonized photons intercepted by the disc, leading to a decrease of the reflection flux \citep{Plant2014}. The balance between the increased reflection flux caused by the disc moving inwards and the decreased reflection flux caused by the increased coronal height could yield a constant reflection flux.

We also find a change in the shape of the QPO rms spectra when the QPO frequency is $\sim1.7$ Hz (Figure \ref{fig:rms_energy}): at low frequencies, the QPO rms first decreases with energy and then becomes flat, while at high frequencies the rms increases with energy. This is similar to what is observed in XTE J1550--564 and XTE J1650--500 \citep{Rodriguez2004,Gierlinski2005}. A possible way to explain this change is an increase in the contribution of the component that dominates the variability in the high-energy band. Previous results show that variability is mainly associated with the hard component \citep[e.g.,][]{Gilfanov2010}, however it is unclear whether the reflection component contributes to the rms spectra. For instance, \citet{Gierlinski2005} found that the reflection variability is also due to variability in the direct emission rather than to changes in the properties of the reflector.

It is worth noting that the phase-lag behavior in GRS 1915+105 also shows an interesting change at QPO frequency around $\sim2$ Hz: below $\sim2$ Hz, the phase lags are positive and increase with energy, whereas above $\sim2$ Hz, the phase lags are negative and decrease with energy \citep{Reig2000,Qu2010,Pahari2013}. GRS 1915+105 has a very different spectral evolution and timing properties compared to GX 339--4, implying that the two systems may differ in the geometry of the accretion flow. The change seen in the phase lags of both system at a very similar QPO frequency makes this an interesting frequency. \citet{Nobili2000} proposed a simple Comptonization model to account for the observed phase lags in GRS 1915+105, assuming that the optical depth of the corona increases as the inner disc radius moves inward. Recently, \citet{Eijnden2016} proposed a differential precession model to explain the phase lags as a function of both energy and frequency. In their scenario, the frequency of $\sim2$ Hz corresponds to a point where the outer-part and the inner-part flows have the same spectral shape. Below $\sim2$ Hz, the inner-part flow has a softer spectrum than the outer-part flow, while above $\sim2$ Hz, it is the opposite. However, we note that these models cannot explain the phase-lag behavior in GX 339--4 due to the different energy-dependence of the QPO lag. In GX 339--4, the QPO phase lags could originate from different mechanisms. At high QPO frequencies, the reflection process should have a significant contribution to the phase lags to produce the observed lag--energy spectra, while at low QPO frequencies, the reflection process is unnecessary and these lags might be dominated by Comptonization process.

It is difficult to explain the complex phase-lag behavior at the second harmonic and subharmonic observed in GX 339--4. In the systematic analysis of several sources by \citet{Eijnden2017}, the phase lags at the second harmonic remain remarkably constant as a function of the second-harmonic frequency. In the case of GX 339--4, however, we find that both the second-harmonic and subharmonic lags change with frequency during an outburst, despite the fact that the fundamental lags during this period remain more or less constant. The similarity between the evolution of the second-harmonic and subharmonic phase lags suggests that these lags might originate from the same mechanism that differs from the fundamental. Using frequency-resolved spectroscopy, \citet{Axelsson2016} found that the second harmonic spectrum can be described by an additional soft Comptonization component.

\section{Conclusions}

In this paper, we present the evolution of the phase lags associated with the type-C QPO in GX 339--4 during the rising phase of the 2006/2007 outburst. We find that the phase lags at the type-C QPO fundamental show very different behavior when the QPO frequency is below and above $\sim1.7$ Hz. The evolution of the QPO phase lags is very similar to that of the reflection flux. A clear break is present in both the phase lags and reflection flux as a function of QPO frequency. Above the break, the QPO phase lags remain more or less constant. The constant reflection flux during these observations and the reflected feature seen in the lag--energy spectra provide strong evidence that in these observations the QPO phase lags are dominated by the reflection process. The lack of a reflected feature in the lag--energy spectra and the completely different phase lag behavior for the QPOs below $\sim1.7$ Hz imply that another mechanism may produce the lags in these observations. The break in the relation between the QPO lags and the QPO frequency would correspond to a change in the dominant mechanism of the phase lags. We explain the increase of the reflection flux below the break as the moving inward of the accretion disc. The constant reflection flux when the QPO frequency is above $\sim1.7$ Hz might be due to a combination of the change of the inner disc radius and the coronal height.

We also find that when the QPO frequency is below $\sim1.7$ Hz, the QPO rms fractional amplitude decreases with energy, when the QPO frequency is close to $\sim1.7$ Hz, the rms spectrum is almost flat, while when the QPO frequency is above $\sim1.7$ Hz, the QPO rms increases with energy. The evolution of the shape of the rms spectra suggests that the QPO is associated with the hard component.

During the transition from the LHS to the HIMS, which takes place at a QPO frequency of $\sim0.54$ Hz, the second harmonic and subharmonic appear in the PDS. The second-harmonic and subharmonic phase lags show a similar evolution as a function of their centroid frequencies. The lag-energy spectra of the second harmonic show a change from increasing with energy to flat at a second-harmonic frequency of $\sim6$ Hz, while the phase lags at the subharmonic always increase with energy.

\acknowledgments

We are very grateful to Daniel Plant for providing us with the reflection data that we used in this paper, and J. van den Eijnden for insightful discussions. We thank the anonymous referee for providing very useful comments that helped us improving our manuscript. LZ acknowledges the Kapteyn Astronomical Institute for their hospitality during his visit at the University of Groningen. DA acknowledges support from the Royal Society. This research has made use of data obtained from the High Energy Astrophysics Science Archive Research Center (HEASARC), provided by the NASA/Goddard Space Flight Center. This research is supported by the National Program on Key Research and Development Project (Grant No.~2016YFA0400801), the NSFC project 11673023 and the Fundamental Research Funds for the Central University.


\vspace{5mm}



\newpage
\begin{center}
\begin{deluxetable*}{ccccccccc}
\tablecaption{RXTE Observations of GX 339-4 used in this analysis, with the centroid frequencies and phase lags of the type-C QPO and its subharmonic and second harmonic components.\label{tab:log}}
\tabletypesize{\scriptsize}
\tablewidth{0pt}
\tablehead{
\colhead{}    & \colhead{}   & \colhead{} & \multicolumn{2}{c}{Fundamental} & \multicolumn{2}{c}{Second Harmonic} & \multicolumn{2}{c}{Subharmonic} \\
\cline{4-5} \cline{6-7} \cline{8-9}
\colhead{No.} & \colhead{Observation ID} & \colhead{MJD} & \colhead{$\nu_{0}$} & \colhead{Phase Lag} & \colhead{$\nu_{0}$} & \colhead{Phase Lag} & \colhead{$\nu_{0}$} & \colhead{Phase Lag}\\
\colhead{} & \colhead{} & \colhead{} & \colhead{(Hz)} & \colhead{(rad)} & \colhead{(Hz)} & \colhead{(rad)} & \colhead{(Hz)} & \colhead{(rad)}
}
\startdata
1  & 92035-01-01-01 & 54128.94 & $0.14\pm0.01$ & $0.020\pm0.011$ &   \nodata      & \nodata & \nodata & \nodata \\
2  & 92035-01-01-03 & 54130.13 & $0.17\pm0.01$ & $0.028\pm0.012$ &   \nodata      & \nodata & \nodata & \nodata \\
3  & 92035-01-01-02 & 54131.11 & $0.18\pm0.01$ & $0.027\pm0.012$ &   \nodata      & \nodata & \nodata & \nodata \\
4  & 92035-01-01-04 & 54132.09 & $0.20\pm0.01$ & $0.013\pm0.014$ &   \nodata      & \nodata & \nodata & \nodata \\
5  & 92035-01-02-00 & 54133.00 & $0.23\pm0.01$ & $0.025\pm0.009$ &   \nodata      & \nodata & \nodata & \nodata \\
6  & 92035-01-02-01 & 54133.92 & $0.26\pm0.01$ & $0.032\pm0.010$ &   \nodata      & \nodata & \nodata & \nodata \\
7  & 92035-01-02-02 & 54135.03 & $0.29\pm0.01$ & $0.039\pm0.009$ &   \nodata      & \nodata & \nodata & \nodata \\
8  & 92035-01-02-03 & 54136.02 & $0.36\pm0.01$ & $0.016\pm0.009$ &   \nodata      & \nodata & \nodata & \nodata \\
9  & 92035-01-02-04 & 54137.00 & $0.43\pm0.01$ & $0.045\pm0.008$ &   \nodata      & \nodata & \nodata & \nodata \\
10 & 92035-01-02-08 & 54137.85 & $0.54\pm0.01$ & $0.036\pm0.021$ &   \nodata      & \nodata & \nodata & \nodata \\
\hline
11 & 92035-01-02-07 & 54138.83 & $0.89\pm0.01$ & $0.094\pm0.010$ &   \nodata      & \nodata         & $0.45\pm0.02$ & $0.113\pm0.010$ \\
12 & 92035-01-02-06 & 54139.94 & $0.99\pm0.01$ & $0.108\pm0.018$ & $1.99\pm0.01$  & $0.251\pm0.010$ & $0.76\pm0.01$ & $0.163\pm0.021$ \\
13 & 92035-01-03-00 & 54140.20 & $1.13\pm0.01$ & $0.111\pm0.028$ & $2.27\pm0.01$  &
$0.299\pm0.018$ &     \nodata   &     \nodata     \\
14 & 92035-01-03-01 & 54141.06 & $1.68\pm0.01$ & $0.167\pm0.018$ & $3.34\pm0.01$  &
$0.463\pm0.019$ & $1.14\pm0.03$ & $0.166\pm0.015$ \\
15 & 92035-01-03-02 & 54142.04 & $2.45\pm0.01$ & $0.122\pm0.014$ & $4.93\pm0.02$  &
$0.485\pm0.027$ & $1.52\pm0.05$ & $0.107\pm0.010$ \\
16 & 92035-01-03-03 & 54143.02 & $3.52\pm0.01$ & $0.144\pm0.014$ & $7.11\pm0.03$  &
$0.203\pm0.034$ & $2.13\pm0.09$ & $0.054\pm0.018$ \\
17 & 92428-01-04-00 & 54143.87 & $4.34\pm0.01$ & $0.178\pm0.015$ & $8.76\pm0.03$  &
$0.130\pm0.034$ & $2.15\pm0.03$ & $0.117\pm0.022$ \\
18 & 92428-01-04-01 & 54143.95 & $4.24\pm0.01$ & $0.148\pm0.017$ & $8.49\pm0.04$  &
$0.134\pm0.032$ & $2.20\pm0.06$ & $0.088\pm0.021$ \\
19 & 92428-01-04-02 & 54144.09 & $4.14\pm0.01$ & $0.186\pm0.018$ & $8.37\pm0.04$  &
$0.163\pm0.040$ & $2.16\pm0.06$ & $0.071\pm0.027$ \\
20 & 92428-01-04-03 & 54144.87 & $4.98\pm0.01$ & $0.199\pm0.022$ & $10.03\pm0.05$ &
$0.146\pm0.057$ & $2.49\pm0.04$ & $0.196\pm0.035$ \\
21 & 92035-01-03-05 & 54145.11 & $5.80\pm0.02$ & $0.173\pm0.016$ & $11.96\pm0.17$ &
$0.273\pm0.042$ & $2.93\pm0.04$ & $0.174\pm0.026$ \\
22 & 92428-01-04-04 & 54145.97 & $5.61\pm0.02$ & $0.121\pm0.025$ & $11.34\pm0.16$ &
$0.179\pm0.067$ & $2.87\pm0.07$ & $0.201\pm0.048$ \\
23 & 92035-01-03-06 & 54146.03 & $5.55\pm0.01$ & $0.144\pm0.016$ & $11.34\pm0.17$ &
$0.079\pm0.039$ & $2.80\pm0.06$ & $0.172\pm0.021$ \\
\enddata
\end{deluxetable*}
\end{center}

\end{document}